\documentclass[12pt]{article}
\usepackage{amssymb}
\usepackage{graphicx}
\begin{document}

\title{On  Cosmological Implications of Gravitational Trace Anomaly}
\author{
Neven Bili\'c,
Branko Guberina,
Raul Horvat,\\
Hrvoje Nikoli\'c,
and Hrvoje \v Stefan\v ci\'c
 \\
Rudjer Bo\v skovi\'c Institute\thanks{Electronic mail:
bilic@thphys.irb.hr, guberina@thphys.irb.hr, horvat@lei3.irb.hr,
hrvoje@thphys.irb.hr, shrvoje@thphys.irb.hr},
POB 180, HR-10002 Zagreb, Croatia 
}

\maketitle

\begin{abstract}
We study the infrared effective theory of gravity that stems from the  quantum trace 
anomaly. Quantum fluctuations of the metric induce running of the cosmological constant
and the Newton constant at cosmological scales. By imposing  the generalized
Bianchi identity we obtain a prediction for the scale dependence of the dark matter and 
dark energy densities in terms of the parameters of the underlying conformal theory.
For certain values of the model parameters the dark energy equation of state and the observed
spectral index of the primordial  density fluctuations  can be simultaneously reproduced.
\end{abstract}



\section{Introduction}

The evidence of the cosmic acceleration presumably driven by
dark energy (DE) with negative pressure \cite{Perlmutter} and
precise measurements of the cosmic microwave background radiation
\cite{Bennett,spe} have triggered a renewed interest in the cosmological constant
\cite{Nobbenhuis}.
Among different approaches to study the cosmological constant
a convenient  framework is  based on the effective field theory 
\cite{Weinberg,Donoghue}. This theory is a long distance realization of
quantum gravity reduced to the general theory of relativity supplemented by
the quantum field theory in curved space \cite{Birrel,Buchbinder}.

The cosmological constant $\Lambda$ and the  Newton constant $G$, being parameters 
in the Einstein-Hilbert action,
receive contributions from quantum loops and become  running constants, i.e.,
functions of the running scale \cite{Shapiro,Babic,shap}.
Unfortunately, the running scale, intuitively expected to be of the order of typical
momenta of the particles in loops, cannot be fixed
unambiguously.

It has been shown \cite{ant1} that a consistent infrared modification 
of gravity due to the quantum  trace anomaly implies 
the presence of additional terms in the low energy effective action.
Quantum fluctuations of the metric modify
the classical metric description of general
relativity at cosmological scales
 and provide a mechanism for a screening of the cosmological constant
and the inverse Newton constant.

The effective theory of the conformal factor
induced by the quantum trace anomaly has a non-trivial infra-red (IR) dynamics owing to
the existence of a non-trivial IR stable fixed point \cite{ant6}.
A number of issues based on this idea were addressed in 
the literature\footnote{For recent reviews, see \cite{ant1,shap2}.}.
A formulation if IR quantum gravity in curved space was suggested
\cite{odi1} and 
the logarithmic corrections to scaling relations in the IR regime were studied
\cite{odi2}.
The IR dynamics of the conformal factor was also
investigated in four-dimensional quantum gravity with torsion
\cite{odi3} and a possible curvature induced
phase transitions in IR quantum gravity was suggested
 \cite{odi4}.

In this paper we  
 investigate the
effective low energy gravity supplemented by the Bianchi identity constraint
and its implications for cosmology. We obtain a model in which,
besides a scale dependent cosmological constant,
dark matter (DM) has a noncanonical cosmological scale
dependence. We find that, under reasonable assumptions in the cosmological context,
the running of the DM particle mass is phenomenologically acceptable.
We demonstrate that a reasonable cosmology is obtained for a 
range of parameters required to fit the observed spectral index of the primordial 
density fluctuations.

We organize the paper as follows. In section \ref{anomaly} we
briefly discuss the large scale effects of the trace anomaly.
In section \ref{cosmo} we introduce a cosmological model based on
a generalized Bianchi identity. Finally,
in section \ref{conclude} we summarize our results and give concluding remarks.

\section{Trace anomaly and effective low energy gravity}
\label{anomaly}
In this section we summarize the basic ideas and results
of Antoniadis, Mazur, and Mottola
\cite{ant2} concerning the gravitational trace anomaly.
The effects of the trace anomaly in the conformal sector of gravity may be
studied by making use of the conformal parameterization
\begin{equation}
g_{\mu\nu}(x) = e^{2 \sigma(x)} \bar g_{\mu\nu}(x),
\label{eq101}
\end{equation}
where $\bar g_{\mu\nu}$ is a fixed fiducial metric.
The total low energy effective action is 
\begin{equation}
S=S_{\rm EH}+S_{\rm matt}+S_{\rm anom},
\label{eq102}
\end{equation}
where
\begin{equation}
S_{\rm EH}=\frac{1}{16\pi G} S_2+\frac{\Lambda}{8\pi G} S_0=
\frac{1}{16\pi G}\int d^4x \sqrt{-g}(R-2\Lambda)
\label{eq103}
\end{equation}
is the classical Einstein-Hilbert action,
$S_{\rm matt}$ is the action that contains the matter fields
and $S_{\rm anom}$ is the anomaly induced effective 
action \cite{ant6,rie,ant7,ant5}

\begin{equation}
S_{\rm anom}=\int d^4x \sqrt{-g}\left[bF\sigma
+b'\left(E-\frac{2}{3}\Box R\right)\sigma +2b'\sigma\Delta_4\sigma\right] .
\label{eq104}
\end{equation}
The differential operator $\Delta_4$ defined as
\begin{equation}
\Delta_4 \equiv \Box^2 + 2 R^{\mu\nu}\nabla_\mu\nabla_\nu - \frac{2}{3} R \Box +
\frac{1}{3}(\nabla^\mu R)\nabla_\mu \,,
\label{eq105}
\end{equation}
is the unique conformally invariant 4-th order operator.
The parameters $b$ and $b'$ are the coefficient that multiply
respectively the square of the Weyl tensor
\begin{equation}
F\equiv C_{\mu\nu\rho\lambda}C^{\mu\nu\rho\lambda}= R_{\mu\nu\rho\lambda}R^{\mu\nu\rho\lambda}-2R_{\mu\nu}R^{\mu\nu}
+\frac{1}{3}R^2
\label{eq1106}
\end{equation}
 and the Gauss-Bonnet term
\begin{equation}
E=R_{\mu\nu\rho\lambda}R^{\mu\nu\rho\lambda}-4R_{\mu\nu}R^{\mu\nu}
+R^2
\label{eq106}
\end{equation}
that appear in the most general expression for the 4-dimensional 
trace anomaly \cite{duff}.
These parameters depend on the matter content of the theory coupled to
$\sigma$.
If only the contribution of free massles fields are taken into account,
$b$ and $b'$ take on the values \cite{ant7}
\begin{equation}
b= -\frac{1}{16\pi^2}\frac{1}{120}(N_S+3N_F+12N_V-8)
 +b_{\rm grav}
\label{eq107b}
\end{equation}
\begin{equation}
b'=-\frac{1}{32\pi^2}Q^2= -\frac{1}{16\pi^2}\frac{1}{360}(N_S+\frac{11}{2}N_F+62N_V-28) 
+b'_{\rm grav}
\label{eq107}
\end{equation}
where $N_S$, $N_F$, and $N_V$ are the numbers
of scalar, Weyl fermion and vector fields, respectively.
The integers
$-8$ and $-28$ come from the
spin-0  metric and ghost fields while $b_{\rm grav}$ and $b'_{\rm grav}$ 
are the contributions of the spin-2 metric fields.
In the following we treat $Q^2$ as a free parameter since
the contributions
beyond the Standard Model are not well known and
the spin-2 contributions in (\ref{eq107b}) and (\ref{eq107})
  depend 
on the model of quantum gravity and are still an open
issue.  The one loop calculations 
in the Einstein theory and in the Weyl-squared theory
give similar results \cite{ant7}. 
However, it is not obvious that
the Weyl-squared Lagrangian is  appropriate to account for 
the contribution of traceless graviton degrees of freedom
and the Einstein theory is plagued with several well-known difficulties.
Although it follows from  (\ref{eq107})
 that $Q^2>0$ for all free matter fields,
we allow $Q^2$ to take on negative values.
Negative sign contributions can
be obtained, e.g., in some
extended models of
conformal supergravity \cite{fra2}
(for additional references, see also \cite{shap2}).

The scale invariant
effective theory that gives rise to (\ref{eq104}) 
has a non-trivial IR  dynamics owing to
the existence of a non-trivial IR stable fixed point \cite{ant6}.
The scale invariance in this theory
persists even at the quantum level.
The sectors
of a theory, the scale invariance of which persists at the quantum level, have
 recently been dubbed ``unparticle stuff" \cite{geo}.
 These sectors, if coupled  to the
standard model sector, seem to cause novel observable effects which could 
perhaps be detected in the future
experiments at TeV scale.
In particular, one could also expect that the
unparticle stuff gives additional contributions to $Q^2$.

The quantum fluctuations of the conformal factor are responsible
for a screening of the cosmological and
inverse Newton coupling constants \cite{ant1,ant2}.
The anomalous dimension of an operator with the classical conformal weight $\omega$
is given by the quadratic equation \cite{ant2,ant3}
 \begin{equation}
\beta_p = 4-p+\frac{\beta_p^2}{2Q^2}\, ,
\label{eq108}
\end{equation}
with the solution
\begin{equation}
\beta_p = Q^2 -\sqrt{Q^4-(8-2p)Q^2} \, ,
\label{eq108b}
\end{equation}
where $p=4-\omega$ is the classical conformal codimension.
The full scaling dimension $\Delta_p$ is related to the classical dimension
by \cite{ant4}
 \begin{equation}
\Delta_p= 4\left( 1-\frac{\beta_p}{\beta_0}\right).
\label{eq108a}
\end{equation}
The operators $S_0$ and $S_2$ appearing in
(\ref{eq103}) aquire anomalous dimensions $\beta_0$ and $\beta_2$,
respectively, and scale with volume according to
\begin{equation}
S_0\sim V ; \hspace{1cm} S_2 \sim V^{\beta_2/\beta_0} ,
\label{eq109}
\end{equation}
whereas the corresponding couplings 
scale inversely 
\begin{equation}
\frac{\Lambda}{8\pi G}\sim V^{-1} ; \hspace{1cm}
\frac{1}{16\pi G} \sim V^{-\beta_2/\beta_0}.
\label{eq109b}
\end{equation}
By similar considerations one finds the scaling laws for fermion
 and boson masses
\begin{equation}
m_F\sim V^{-\beta_3/\beta_0};
 \hspace{1cm}
m_B^2 \sim V^{-\beta_2/\beta_0}.
\label{eq109c}
\end{equation}

Another effect of the quantum fluctuations of the conformal factor concerns
the departure of the fractal space-time dimension from its classical value
\cite{ant2}.
It turns out that the volume $V$ does not scale with geodesic distance $l$ naively as 
$V\sim l^4$. Rather, it scales according to 
\begin{equation}
V\sim l^{d_{\rm H}} \, ,
\label{eq117}
\end{equation}
where $d_{\rm H}$ is the Hausdorff dimension which classically equals 4.
The calculation based on the quantum gravity distance defined by the heat kernel of the
operator $\Delta_4$ yields the Hausdorff dimension expressed in terms of the 
parameter $Q^2$ \cite{ant2}
\begin{equation}
d_{\rm H}=-4\frac{\beta_8}{\beta_0}.
\label{eq113}
\end{equation}
where $\beta_8$ and $\beta_0$ are given by 
(\ref{eq108b}).
In the course of the cosmological evolution
the physical geodesic distance $l$ scales as
$l \sim a$ and hence the volume scales with $a$ as
\begin{equation}
V \sim a^{d_{\rm H}}.
\label{eq118}
\end{equation}
As has been emphasized in \cite{ant2},
the scaling
relations (\ref{eq109b}) and  (\ref{eq109c}) 
of dimensionful quantities 
are not directly physically relevant since the units in which the volume
is measured have not been specified.
A physically meaningful scaling relation is obtained
when a product of powers of two quantities is formed
so that its naive dimension is zero.
In this way one of the quantities is measured in units of the other.

Combining (\ref{eq109b}), (\ref{eq109c}), and (\ref{eq118}),
the net effect is a cosmological scale 
dependence of  the dimensionless quantities  \cite{ant2}
\begin{equation}
G\Lambda=G_0\Lambda_0 a^\mu,
\label{eq110}
\end{equation}
\begin{equation}
Gm_F^2=G_0 m_{F0}^2 a^\nu ,
\label{eq111}
\end{equation}
\begin{equation}
Gm_B^2=G_0 m_{B0}^2 a^0 ,
\label{eq111a}
\end{equation}
where the exponents $\mu$ and $\nu$ are given by
\begin{equation}
\mu=d_{\rm H}\left(2\frac{\beta_2}{\beta_0}-1\right);
\hspace{1cm} 
\nu=d_{\rm H}\frac{\beta_2-2\beta_3}{\beta_0}.
\label{eq114}
\end{equation}
In the limit of large $Q^2$ one finds
\begin{equation}
d_{\rm H}\simeq 4\left(1+\frac{4}{Q^2}\right) ;
\hspace{1cm}
\mu \simeq -\frac{4}{Q^2};
\hspace{1cm} 
\nu \simeq \frac{1}{Q^2}.
\label{eq115}
\end{equation}
Thus, for positive $Q^2$
 the cosmological constant decreases, 
the fermion masses increase, and the
boson masses remain constant with increasing cosmological scale $a$,
when these quantities are measured in units of the Planck mass.

\section{Cosmology by the generalized Bianchi identity}
\label{cosmo}
A scale-setting procedure based on the implementation of the generalized Bianchi
identity was established  and 
successively applied \cite{Babic2} both to the 
effective field theory of gravity and matter, and to the nonperturbative
quantum gravity \cite{Reuter}. Besides, it was found 
\cite{Horvat,GuberinaJCAP} that both theories are
consistent with holographic dark energy \cite{Cohen}, provided the running
scale  was identified with an infrared cutoff roughly equal to the inverse
size of the system.

It seems reasonable to assume that
the quantum effects of the conformal factor 
which we have discussed in section \ref{anomaly}
do not alter the Einstein field equations at large distances,
apart from the gravitational dressing of $G$ and $\Lambda$
due to these effects.
Then, the contracted  Bianchi identity of the Einstein tensor yields the
 conservation law
\begin{equation}
 \nabla^{\mu} \left[G (T_{\mu \nu} + g_{\mu \nu} \rho_{\Lambda}) \right]=0 \, ,
\label{eq201}
\end{equation}
where
 the energy-momentum tensor
 takes
the usual perfect fluid
form 
$T_{\mu\nu}=(p+\rho)u_\mu u_\nu +p g_{\mu\nu}$.
In the comoving frame with FRW metric, equation (\ref{eq201}) becomes
\begin{equation}
a\frac{d}{da}\,\left[G(\rho+\rho_{\Lambda})\right]+3G(p+\rho)=0\,.
\label{eq203}
\end{equation}
where the cosmological scale $a$ satisfies the Friedmann equation in
flat space
\begin{equation}
H^2\equiv \left(\frac{\dot{a}}{a}\right)^2=\frac{8\pi G}{3}(\rho+\rho_\Lambda) \, .
\label{eq204}
\end{equation}
In (\ref{eq203}) we implicitly assume scale dependent
$\rho_{\Lambda}$ and $G$ and  the scale dependence of the
DM density $\rho$ need not  be canonical $a^{-3-3w}$.
Furthermore, we assume that matter is nonrelativistic, i.e.,
 that the matter energy density can be written as $\rho=\sum_i n_i m_i$, 
where $n_i$ are the particle number 
densities and $m_i$ are the masses of the particle species. 
If nonrelativistic matter consists of $l$ fermionic and $k$ bosonic species,
 we have $l+k+2$ equations (equations (\ref{eq110})-(\ref{eq111a}) and (\ref{eq203})) 
for $2+2l+2k$ quantities.
Although we do not expect $k$ and $l$ to be much larger than 1, additional assumptions 
are needed in order to solve the equations uniquely.

Postulating (\ref{eq203}), the scale behavior of $\rho$ depends
on the scaling of ${\Lambda}$ and $G$.
Equation (\ref{eq110}) gives the scaling of the dimesionless product
${\Lambda}G$ and does not determine the scaling of the dimensionful
parameters $G$ and ${\Lambda}$
separately.
However, we are allowed to choose the units of measurement such
that one chosen dimensionful parameter is fixed.
We may, e.g., investigate  three obvious choices when
one of the three quantities
 $G$, $\Lambda$, or $\rho_\Lambda$ is fixed.

A more general case that includes the above mentioned choices is 
obtained if we
alow $G$ and $\Lambda$ to vary as  powers of $a$ restricted only by equation (\ref{eq110}).
Hence,  we set
\begin{equation}
G=G_0 a^{\alpha}; \hspace{1cm} \Lambda=\Lambda_0 a^{\mu-\alpha} ,
  \label{eq206}
\end{equation}
where $\alpha$ is an arbitrary parameter.
With this ansatz, $\rho$ may be expressed in terms
of the fixed dimensionful parameter $\Lambda^\alpha/G^{\mu-\alpha}=\Lambda_0^\alpha/G_0^{\mu-\alpha}$
\begin{equation}
\rho(a)=\left(\frac{\Lambda^\alpha}{G^{\mu-\alpha}}\right)^{2/\mu} \rho_*(a)
=\Omega_\Lambda \rho_c f(a) ,
  \label{eq207}
\end{equation}
where $f(a)=
{8\pi}(G_0\Lambda_0)^{2\alpha/\mu-1}\rho_*(a)$
 is a dimensionless function of $a$, $\rho_c$ is
the critical density at present, and the constant
$\Omega_\Lambda$ which  may be fixed from observations
is of the order
of the present fraction of DE density.

Plugging (\ref{eq206}) and (\ref{eq207}) in (\ref{eq203}) and 
neglecting the DM pressure we obtain a differential
equation for
$f$
\begin{equation}
a\frac{df}{da}+(3+\alpha)f+
(\mu-\alpha) a^{\mu-2\alpha}=0,
 \label{eq209}
\end{equation}
with the solution
\begin{equation}
f=C a^{-3-\alpha}-\frac{\mu-\alpha}{3+\mu-\alpha}a^{\mu-2\alpha}.
  \label{eq210}
\end{equation}
The integration constant $C$ is for small 
$\mu$ and $\alpha$  roughly 
$C\simeq (1-\Omega_\Lambda)/\Omega_\Lambda$
so that $\rho$  fits the present DM density.
Although the parameter $\alpha$ is arbitrary, the small values
 $|\mu|\ll 1$ and $|\alpha|\ll 1$ are phenomenologically desirable
since the variation of $\Lambda$ and $G$ should not be too large to spoil
the observational
constraints. 
Hence, equation (\ref{eq210}) implies  a slight modification
of the DM density. However, it may be easily seen that the effective 
DM density in the Friedmann equations remains canonical.
Using (\ref{eq206}), (\ref{eq207}), and (\ref{eq210})  the Friedmann equations
may be written in the form
\begin{equation}
\left(\frac{\dot{a}}{a}\right)^2=\frac{8\pi G_0}{3}(\rho_{\rm DM}+\rho_{\rm DE}) \, ,
\label{eq211}
\end{equation}
\begin{equation}
\frac{\ddot{a}}{a}=-\frac{4\pi G_0}{3}(\rho_{\rm DM}+\rho_{\rm DE}+3p_{\rm DE}) \, ,
\label{eq212}
\end{equation}
where the effective DM and DE densities are given by
\begin{equation}
\rho_{\rm DM}=(1-\Omega_{\rm DE})\rho_c a^{-3},
  \label{eq213}
\end{equation}
\begin{equation}
\rho_{\rm DE}=\Omega_{\rm DE}\rho_c
a^{\mu-\alpha},
  \label{eq214}
\end{equation}
and the effective DE pressure is 
\begin{equation}
p_{\rm DE}=-\left(1+\frac{\mu-\alpha}{3}\right)\rho_{\rm DE}\, .
  \label{eq215}
\end{equation}
In (\ref{eq213}) and (\ref{eq214}) we have introduced the constant
$\Omega_{\rm DE}$ which we identify with the present fraction of DE.
In this way we fix
the arbitrary
constants $\Omega_{\Lambda}$ and $C$ which are related to $\Omega_{\rm DE}$ by
\begin{equation}
\Omega_{\Lambda}= \left(1+\frac{\mu-\alpha}{3}\right) \Omega_{\rm DE}; 
\hspace{1cm}
C=\frac{(1-\Omega_{\rm DE})}{\Omega_{\Lambda}}\, .
  \label{eq216}
\end{equation}
Equations (\ref{eq211})-(\ref{eq215}) show that the models satisfying 
(\ref{eq110}) and (\ref{eq206})
 closely mimic the cosmology with standard cold DM  and dark energy 
 with a constant equation of state (\ref{eq214})
(XCDM cosmologies \cite{tur}), at least at the level of the 
global evolution of the universe.
There is an obvious twofold implication of this result.
First, in the analysis of the observational data, 
especially the SN Ia observations, one should bare  in mind that
a good fit to an XCDM cosmology may be interpreted as
 a signal for the dynamics given by
(\ref{eq206}).
Second,
 the observational constraints on XCDM  cosmologies
 can be
used to estimate
the parameters of the models defined by (\ref{eq110}) and (\ref{eq206}).
The values of $\alpha$ equal to 0, $\mu$, and $\mu/2$ correspond to
fixed $G$, $\Lambda$, and $\rho_{\Lambda}$, respectively.
Although the simplest result is obtained for $\alpha=\mu$,  the natural choice is $\alpha=0$, i.e., fixed $G$, since a variation of
$G$ is subject to the  most restrictive
observational constraints\footnote{For an extensive
 discussion of models with variable $\Lambda$ and fixed $G$,
see \cite{coo}.}.
In this case we obtain the DE equation of state
\begin{equation}
w=-1-\frac{\mu}{3}\, .
  \label{eq218}
\end{equation}
The observational constraint on the equation of state \cite{spe}
$w=-0.97\pm 0.07$ yields 
\begin{equation}
-0.3 \leq\mu\leq 0.12 \, ,
  \label{eq219}
\end{equation}
which in turn constrains the parameter $Q^2$.
Using (\ref{eq115}) we obtain
\begin{equation}
Q^2 < -33\,  \hspace{0.5cm} {\rm or} \hspace{0.5cm} Q^2 > 13\, ,
  \label{eq222}
\end{equation}
where $Q^2< 0$ would imply a phantom cosmology.

Another constraint on $Q^2$ comes from the primordial density
fluctuations \cite{ant3}.
It is usual to assume that the 
two point correlation function of the
primordial density fluctuations $\delta \rho$,
should behave like 
\begin{equation}
\left\langle \delta \rho(k)\delta \rho(-k)\right\rangle \sim k^n \, ,
\end{equation}
where the exponent $n$, called the {\it spectral index}, need not be constant
over the entire range of wave numbers.
According to Harrison and Zel'dovich
\cite{har} the primordial density
fluctuations  should be characterized by a spectral index $n=1$.
In other words,  the observable giving rise to these
fluctuations has naive scaling dimension $p=2$.
This naive scaling dimension reflects the fact that the density fluctuations 
 are related to the metric fluctuations by Einstein's equations
$\delta R \sim G\delta\rho$ in which the scalar curvature is second order
in derivatives of the metric.
Hence, the two point spatial correlations  
$\left\langle \delta R (x)\delta R (y)\right\rangle$ should behave
as  $|x-y|^{-4}$ or $|k|^1$ in Fourier space.

More generally,
as a consequence of conformal invariance
the two-point correlation function of an observable
$\cal{O}$ with dimension $\Delta$ is given by
\begin{equation}
\left\langle {\cal{O}}(x){\cal{O}}(y)\right\rangle \sim |x-y|^{-2\Delta} \, ,
\end{equation}
at equal times in three dimensional flat space. In $k$-space
this becomes 
\begin{equation}
\left\langle {\cal{O}}(k){\cal{O}}(-k)\right\rangle \sim |k|^{2\Delta-3} \, ,
\end{equation}
Thus, the spectral index of an observable of dimension $\Delta$
is defined by
\begin{equation}
n=2\Delta-3 \, ,
  \label{eq220}
\end{equation}
Hence, the Harrison-Zel'dovich spectral index $n=1$ 
corresponds to the classical dimension $\Delta=2$ of
the primordial density
fluctuation.

If the conformal fixed point behavior \cite{ant6} described in section 
\ref{anomaly} dominates at cosmological scales
then the scaling dimension $\Delta_p$ of an observable with classical dimension $p$
is given by (\ref{eq108a})
as required by the conformally invariant fixed point for gravity.
As a consequence of (\ref{eq220}), with $\Delta_2$
instead of $\Delta=2$,
 a deviation from the Harrison-Zel'dovich spectrum
is obtained.
For large $Q^2$ one finds \cite{ant1,ant3}
\begin{equation}
n\simeq 1+\frac{4}{Q^2} \, .
  \label{eq221}
\end{equation}
A comparison of
$n$ thus calculated  with recent observations
yields a constraint  $|Q^2|>80$.

The  favored  WMAP value seems to be $n=0.95$ which requires
a large negative $Q^2\simeq -80$.
With this value, we obtain dark energy
of the phantom type with
$w\simeq -1.02$, which is consistent with 
SN Ia and WMAP observations.
This result justifies a relaxation 
of the allowed range for the parameter $Q^2$.
The above considerations show that
using a single value of the parameter $Q^2$ in our model
 it is possible to satisfy the observational constraints  
for two essentially unrelated phenomena:
 the present accelerated expansion of the universe (\ref{eq218})
 and the spectral index of primordial density fluctuations (\ref{eq221}). 
The required negative value of $Q^2$ cannot be easily 
accommodated within the framework 
of the present theory,
 but the phenomenological potential of negative $Q^2$
 is a clear incentive to search for new mechanisms which
could bring $Q^2$ into the negative realm.

The ansatz (\ref{eq206}) is not the most general and does not cover all interesting
possibilities.
For example, it does not include
a natural starting assumption  that the total energy density 
of nonrelativistic matter scales canonically with $a$. 
With this assumption, equations (\ref{eq110}) and (\ref{eq203}) fully determine
 the evolution of $\Lambda$ and $G$.
 However, the canonical scaling of the matter energy density,
 $\rho \sim a^{-3}$,
combined  with the scalings (\ref{eq111}) and (\ref{eq111a}), 
implies that the particle number densities $n_i$ no longer vary as $a^{-3}$
and hence, the particle numbers of individual species are not conserved.
 Although the assumption of canonical scaling of the matter density
 yields a mathematically consistent model,
 we find a strong disagreement with observations in a wide parametric range.
 In particular, for negative $\mu$ we obtain a maximal allowed value of the redshift
 of the order of 1 which is clearly not acceptable.

Another interesting model, not covered by
(\ref{eq206}), is obtained if one assumes that all relevant 
particle species are fermions and that
the corresponding particle number densities scale canonically, i.e., $n_i \sim a^{-3}$.
In this case one can express the mass in terms of $G$ and $a$ from (\ref{eq111}),
 $\Lambda$ in terms of $G$ and $a$ from (\ref{eq110}) and solving (\ref{eq203})
 obtain an evolution equation for $G$ as a function of $a$. However, 
numerical solutions of this equation show that the resulting scaling of $G$ with $a$ 
 is too strong to satisfy observational bounds, even for small values of parameters 
$\mu$ and $\nu$.

\section{Conclusion}
\label{conclude}
 We have studied some DE and DM aspects of the low energy effective theory of
gravity.
This theory is a modified general relativity
in which the Einstein-Hilbert action is supplemented with 
the non-local terms induced by the trace anomaly of massless fields.
These nonlocal  terms
do not decouple for scales $E \ll M_{\rm Pl}$ and therefore
become increasingly important for the present
and future time cosmology. The part of the action that stems from
the trace anomaly contains all infrared relevant terms which are
not contained in the local action.

The testing of the theory vs. observational astrophysics and cosmology
is a long term project, which includes the use of $T^{\rm (anom)}_{\mu\nu}$
as a dynamical source for Einstein's equations.
However, we believe that the effective theory with running $G$ and $\Lambda$
supplemented with the generalized Bianchi identity
may be successfully confronted with cosmological observations.

The effective low energy gravity in the conformal sector
predicts a cosmological scale dependence of various
dimensionless quantities (equations (\ref{eq110})-(\ref{eq111a})).
The scale dependence of 
DM and DE densities, being dimensionful quantities,
depends on 
the choice of a fixed dimensionful parameter.
In particular, fixing the Newton constant $G$
 yields a cosmological constant 
scaling with $a$ as $\Lambda\sim a^\mu$ and a noncanonical scaling
of the DM energy density given by (\ref{eq210}) with $\alpha=0$.
The effective DM and DE densites yield
a reasonable cosmology if the exponent $\mu$ is small and restricted by
the constraint (\ref{eq219}). This constraint in turn yields a constraint
(\ref{eq222}) on  $Q^2$ consistent with the observational bounds on the spectral 
index of the primordial density fluctuations.

In our approach the parameter $Q^2$ that appears in the
effective action (\ref{eq104}) induced by the gravitational anomaly
is treated  as a free parameter.
This parameter
 could,
in principle,  be calculated 
 if one had a complete
information on the conformally invariant sector.
Unfortunately this sector is yet not well known so
a precise theoretical prediction for
$Q^2$ remains an open problem.

\section*{Acknowledgments}
This work was supported by the Ministry of Science,
Education and Sport
of the Republic of Croatia under contracts No. 098-0982930-2864 (N.B., B.G.,
H.N., H.\v S.) and No. 098-0982887-2872 (R.H.), and
partially supported through the Agreement between the Astrophysical
Sector, S.I.S.S.A., and the Particle Physics and Cosmology Group, RBI.

\end{document}